\documentclass[aip,reprint]{revtex4-1}
\usepackage[utf8]{inputenc}
\usepackage{amsmath}
\usepackage{natbib}
\usepackage{graphicx}
\begin{document}

%\preprint{}

\title{Liquid crystal nose based on chiral photonic band gap materials: principles of selective response} %Title of paper



\author{P.V. Shibaev}
%\thanks{}
\affiliation{Physics and Engineering Physics Department, Fordham University, NY, USA}

\author{O. Roslyak}
\affiliation{Physics and Engineering Physics Department, Fordham University, NY, USA}
\author{E. Gullatt}
\affiliation{Physics and Engineering Physics Department, Fordham University, NY, USA}
\author{J. Plumitallo}
\affiliation{Physics and Engineering Physics Department, Fordham University, NY, USA}
\author{U. Aparajita}
\affiliation{Science Department, Borough of Manhattan Community College, CUNY, USA}

\date{\today}

\begin{abstract}
Novel liquid crystalline (LC) compositions are suggested and studied as elements of LC-nose. This allows for optical detection of several volatile organic compounds (VOCs). Ethanol, toluene, pyridine and acetic acid were detected by means of colorimetric and spectroscopic techniques during their diffusion inside chiral elements of LC-nose. Selectivity to different VOCs is enhanced by means of components of liquid crystal matrix with different viscosity, affinities to the solvents, and abilities to form hydrogen bonding. 
\end{abstract}


\maketitle %\maketitle must follow title, authors, abstract and \pacs

\section{Introduction}
There is a continuous need for creating novel materials and devices (artificial noses) that can respond to different kinds of pollutants including volatile organic compounds (VOCs) \cite{liu2012survey}. The VOCs of environmental concern are almost all organic solvents (for example, toluene, ethanol, pyridine, acetic acid, etc.). Most of the current compact and inexpensive gas and VOC sensors are based on metal oxides that can change their resistance in different gases \cite{kohl2001function}. However, these sensors have some drawbacks, making them unsuitable for some applications (low selectivity for different VOCs, a need for a permanent power supply, etc). Organic solvents producing VOCs are used in industrial environment, and there are certain limits on the daily exposure of workers to harmful concentrations of these VOCs (these limits are established by governmental agencies) \cite{chang2010health}. The sensors responding to VOCs should measure a cumulative effect of VOCs’ exposure in a certain time period. Sometimes, they should also be installed in locations where the presence of electrical equipment is not desirable. In our view, the passive sensors based on polymers with different sensitivity to organic solvents is a way to address and solve this problem. 
\par
Efforts were undertaken in order to create sensitive and selective sensors based on different types of polymers and polymer compositions, including crystalline polymers \cite{pilla2009molecular}, nanoimprinted polymers \cite{huynh2015molecularly}, wearable polymer compositions and liquid crystals \cite{wang2018liquid,honaker2019elastic}. Very often environmentally sensitive polymers are organized in an array with different sensitivities of elements to specific compounds \cite{fitzgerald2019barcoded}. This type of organization provides higher selectivity of the whole sensor. Increased sensitivity can also be obtained by using photonic band gap materials \cite{banerjee2009enhanced}. Thus, an array of photonic band gap materials may provide both better selectivity and sensitivity to a number of VOCs and is being tested in this publication.
\par
Photonic band gap materials in a form of chiral liquid crystals (CLCs) were used for detecting VOCs  as a single composition  material \cite{kirchner2006functional,han2010optical,mujahid2010solvent,sutarlie2011cholesteric,chang2011optical} for detecting amines \cite{kirchner2006functional}, alcohols and aldehydes \cite{han2010optical,mujahid2010solvent,sutarlie2011cholesteric}.  Chiral liquid crystals can be imagined as nematic layers making a full rotation over a period of the helical pitch. They possess a photonic band gap (selective reflection band, SRB) for circularly polarized light with the same sense of polarization as their rotation. If nematic planes are parallel to the substrate the whole structure displays the SRB centered at a wavelength \cite{belyakov1992diffraction} given by a product of average refractive index and helical pitch: 
\begin{equation}
\label{Eq:1}
    \lambda=n_{av} p
\end{equation}
The general principles of gas detection by CLCs are all based on recording changes (responses) in the spectral position of the SRB and were discussed in \cite{chang2011optical,shibaev2015rebirth}. In these studies the composition of LC sensor was not tailored for detection of wide range of VOCs and the response was studied for just a few VOCs. 
\par
\begin{figure}[ht!]
\centering
\includegraphics[width=\columnwidth]{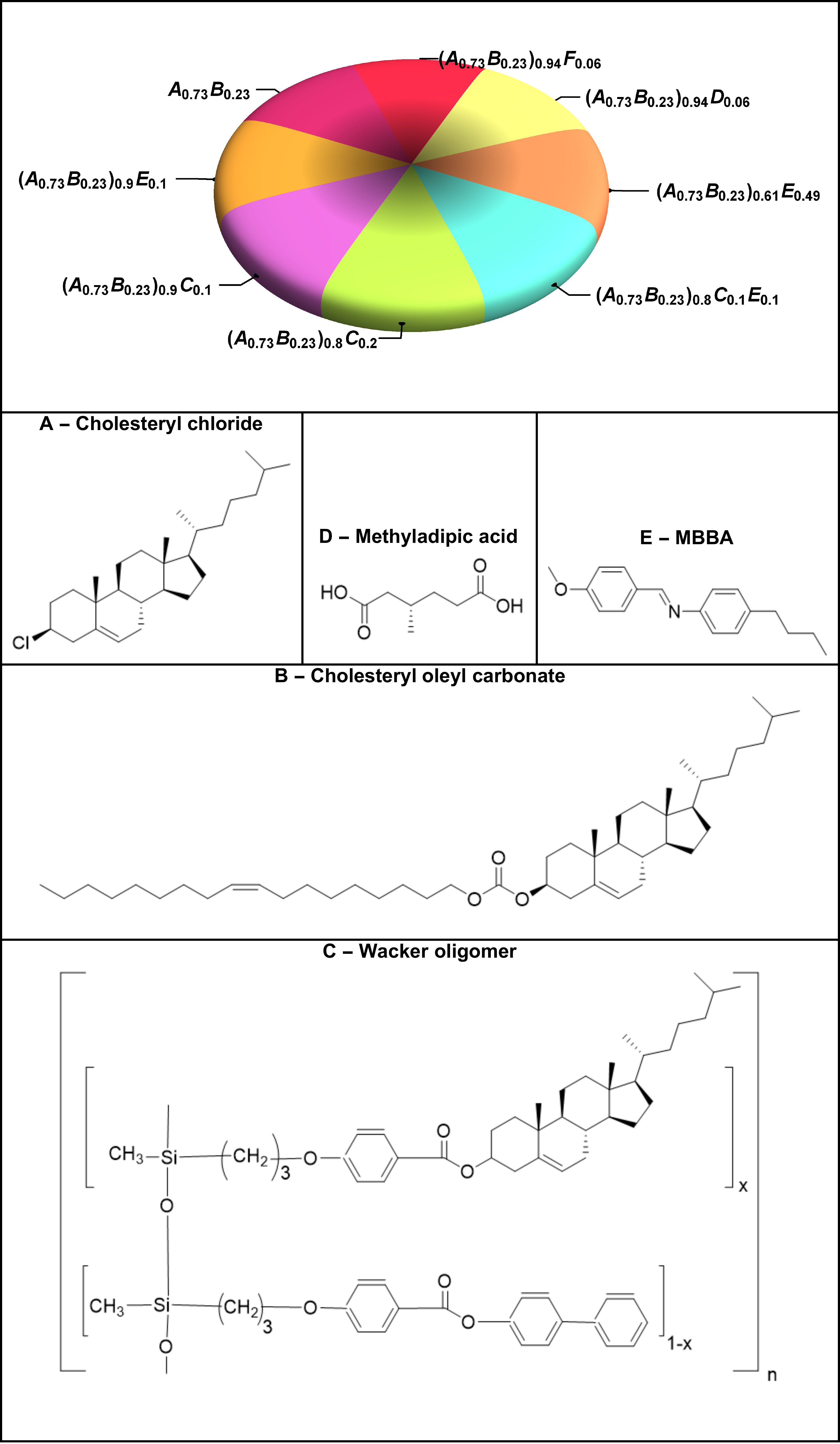}
\caption{The main array component of the proposed LC-nose. Various compositions are shaped into droplets schematically indicated by colors.
The compositions are comprised of compounds (indicated by letters) whose relative concentrations are shown in subscripts. The transmission of un-polarized light through droplets is recorded via CCD camera. The droplets are further subjected to VOC diffusion and changes in the transmission reflect changes in the order parameter and helical pitch of the CLC's.}
\label{FIG:1M}
\end{figure}

Recently \cite{shibaev2019liquid}, a novel approach based on exploring materials with different viscosities as elements of the sensor was suggested and tested for two VOCs ( cyclohexane and ethanol).  In the present publication we suggest a generalized approach and describe a sensor (multielement array) with individual elements composed of CLCs with compositions tailored for detecting specific VOCs (Fig.\ref{FIG:1M}). This sensor displays a selective response to a number of VOCs of different polarities ranging from the acetic acid to toluene and essentially constitutes a novel type of sensor - liquid crystal nose ( LC-nose). Indeed, each element of the array (droplet of CLC with a composition outlined in Fig.\ref{FIG:1M} top panel) is characterized by an individual response rate to a particular VOC, playing a role of nose’s receptor. The whole array then will display a characteristic response pattern for any particular VOC. The pattern is essentially a “fingerprint” of response to certain VOC. 
\par
The approach is based on the following theoretical considerations. 
The width of the SRB is defined by the following equations defining positions of band edges \cite{belyakov1992diffraction}
\begin{equation}
\label{Eq:2}
    n_o p \leq \lambda \leq n_e p
\end{equation}
,where $n_o$ and $n_e$ are ordinary and extraordinary refractive indices along nematic planes.
CLCs are perhaps the most suitable materials for detecting a cumulative effect of exposure to different types of VOCs. Indeed, the diffusion of VOC molecules inside CLC matrix  may result in changes of order parameter $S$, refractive indices $n_e,\; n_o$ and corresponding birefringence $\delta n = n_e- n_o$, helical pitch $p$ and potentially the twisting power of chiral solutes defined as 
\begin{equation}
\label{EQ:Beta}
\beta = \frac{1}{p \times c}    
\end{equation} 
along with concentration of chiral molecules $c$ producing a twisted structure. 
\par
The change of each parameter ($S, \; c, \;  p$ and $\beta$) alone will result in optical response: the shift of the stop band that can be easily detectable by means of spectroscopic or photographic techniques. However, these parameters are not independent as indicated , for instance, by Eq.\eqref{EQ:Beta}. Order parameter and birefringence are also related via
\begin{gather}
    n_e = n_{av} + \frac{2}{3}S\Delta n\\
    \notag
    n_o = n_{av} - \frac{1}{3}S\Delta n
\end{gather}
, where $\Delta n$ is a birefringence in ideally oriented chiral planes ($S=1$).
%--------------
\section{Experiment setup and discussion}
The magnitude and rate of spectral response to the action of VOCs also depend on the diffusion rate of VOC molecules inside the matrix that in turn depends on the chemical nature of diffusant and viscosity of the matrix \cite{shibaev2019liquid}. 
Taking these facts into consideration, we suggest an array-based sensor with elements consisting of multi-component CLCs responding to a particular VOC by changing predominantly either order parameter $S$ or helical pitch $p$. 
This is achieved by mixing components of CLCs with different abilities to form specific physical bonds with the diffusant. 
Also, different affinities to the diffusant increase the selectivity of response. 
The variable viscosities of some elements of the sensor affect the diffusion rate of different VOCs, also contributing to improved selectivity of response. 
The diffusant produced by a solvent that can dissolve a particular component of CLC will also affect the whole composition more strongly, if the concentration of that component is higher. 
The chemical structures of the components are shown in Fig.\ref{FIG:1M}.
\par
Compounds $\texttt{A}$ and $\texttt{B}$ in this figure are derivatives of chiral cholesterol, and their mixture (23:77 by weight) forms liquid crystal at room temperature with the selective reflection band at $620 \; \texttt{nm}$. Both compounds are dissolved in toluene and should be sensitive to its vapor. Compound $\texttt{E}$ is \textit{4-Methoxybenzylidene-4-butylaniline} (MBBA) forming a nematic liquid crystal in a pure form. It is also able to form hydrogen bonds with acids and alcohols through the nitrogen atom presented in its structure. In a pure form it is dissolved in more polar solvents including ethanol and should be sensitive to its vapors. Compound $\texttt{D}$ is methyladipic acid that exists in chiral and racemic (compound $\texttt{F}$) forms. In its chiral form it twists the nematic phase forming left-handed chiral nematic phase. The presence of two acidic groups suggests that it can form strong hydrogen bonds with VOCs of acetic acid and pyridine and weak hydrogen bonds with ethanol. Such a complex of MAA with these compounds should have a different (presumably higher) twisting power than pure chiral MAA and therefore may affect the response by decreasing untwisting of cholesteric helix. Compound $\texttt{C}$ is siloxane based oligomer forming chiral glass at room temperature with the selective reflection band located at $550\; \texttt{nm}$. It is dissolved in non-polar solvents, provides sensitivity to their vapors, and it is also used here to vary the viscosity of the array elements, making them less sensitive to VOCs of non-polar solvents. The structure of the compounds suggests that their mixtures (compositions) presented in Fig.\ref{FIG:1M}(top panel) may respond to different VOCs through the aforementioned mechanisms involving specific inter-molecular interactions. The following four organic VOCs with different polarities and abilities to form hydrogen bonding  were used in experiments: toluene, pyridine, ethanol, acetic acid.
\par
\begin{figure}[ht!]
\centering
\includegraphics[width=\columnwidth]{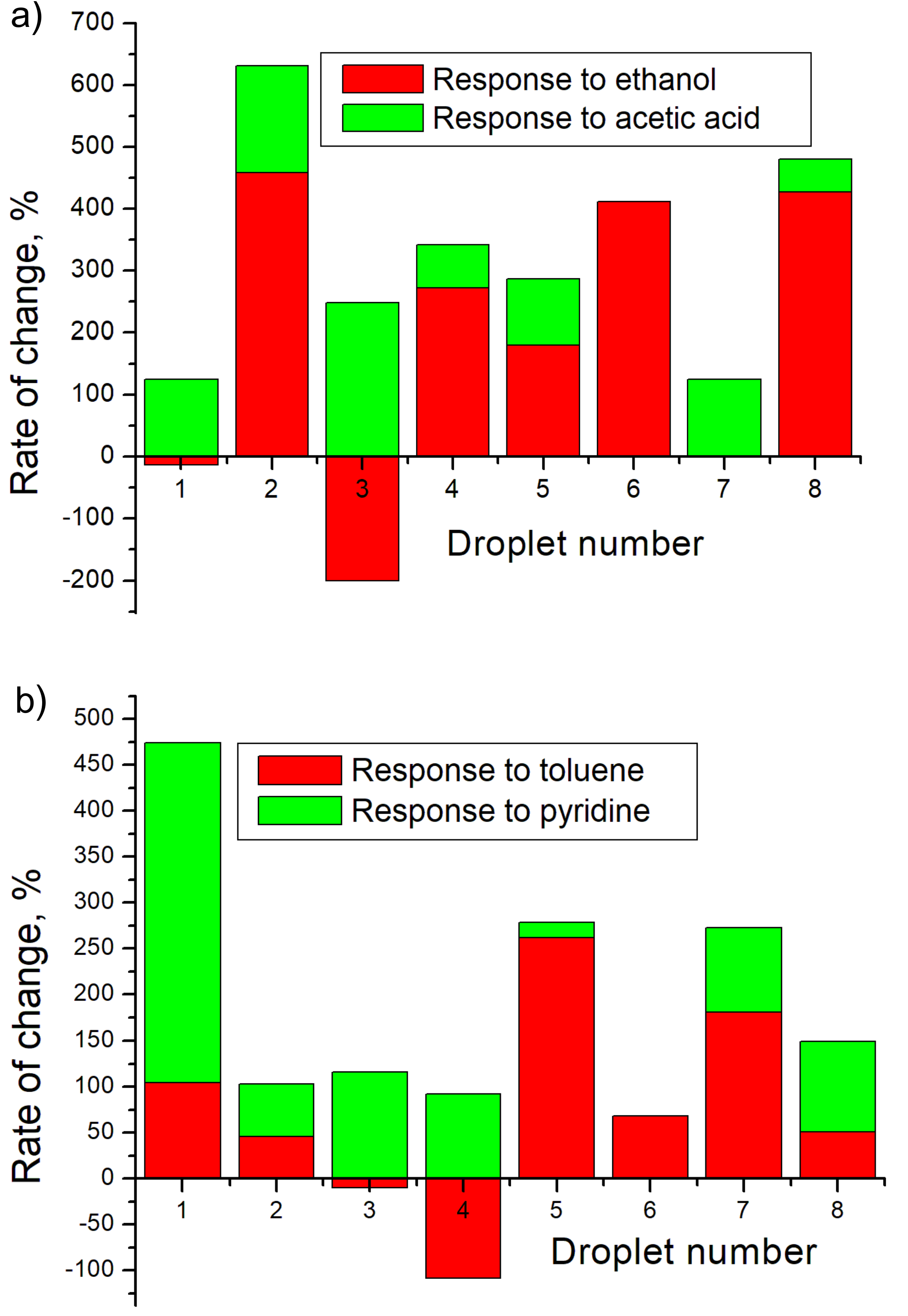}
\caption{Sensor response to different VOCs: a) ethanol and acetic acid, b) toluene and pyridine. The droplets are enumerated from droplet 1 (composition $\texttt{A}_{0.73} \texttt{B}_{0.23}$) counterclockwise as in the top panel of Fig.\ref{FIG:1M}}
\label{FIG:3}
\end{figure}

The sensor array comprised of eight elements shown in the top panel of Fig.\ref{FIG:1M} was placed on a glass substrate with a dark background.  The array was placed in a glassy Petri dish and exposed to a particular VOC freely evaporating from solution at room temperature $21^\circ \texttt{C}$ and reaching saturated concentration inside Petri dish. VOCs diffused inside the droplets that resulted in 1) changes of droplets helical pitch and color, 2) decrease of order parameter and isotropisation.  The VOCs of four solvents ethanol, toluene, pyridine and acetic acid were used and identified by LC-nose in experiments.  The care was taken of creating droplets of approximately the same mass. The color changes were continuously recorded by CCD camera during $30-40$ minutes, followed by the analysis of color intensities in the red and green channels for different time intervals. The response function was then calculated for each element for the areas corresponding to c.a. $20\%$ of the droplet surface near its edge as the ratio of intensity changes in the red and green channels. It can be seen that all responses are different and, therefore, these VOCs can be selectively identified by this sensor. It is important to note that color changes (and a shift of the selective reflection band) happen first, followed by the isotropisation of each droplet. It means that initially some diffusion of VOC molecules occurs inside liquid crystalline phase of the droplet, followed by isotropisation. The sensor displays the fastest response to pyridine and toluene VOCs (isotropisation occurs already after 3-4 minutes of exposure) and the slowest to ethanol for which isotropisation starts in the thinnest part of the droplet 1 only after 40 min of exposure (Fig.\ref{FIG:3} a). The strongest response is displayed by 1st droplet (composition $\texttt{A}_{0.73} \texttt{B}_{0.23}$) to pyridine (Fig. \ref{FIG:3} (b)) and by 2nd droplet to ethanol. Interestingly, the response of composition $\left({\texttt{A}_{0.73} \texttt{B}_{0.23}}\right)_{0.61}\texttt{E}_{0.49}$ (with MBBA) to ethanol is much faster than response of the pure composition $\texttt{A}_{0.73} \texttt{B}_{0.23}$, confirming higher “affinity” between MBBA and ethanol. Composition $\left({\texttt{A}_{0.73} \texttt{B}_{0.23}}\right)_{0.9}\texttt{C}_{0.1}$ with higher concentration of Wacker oligomer (not dissolved in polar solvents) is almost non-responsive to ethanol and responds well to toluene.
\par
\begin{figure}[ht!]
\centering
\includegraphics[width=\columnwidth]{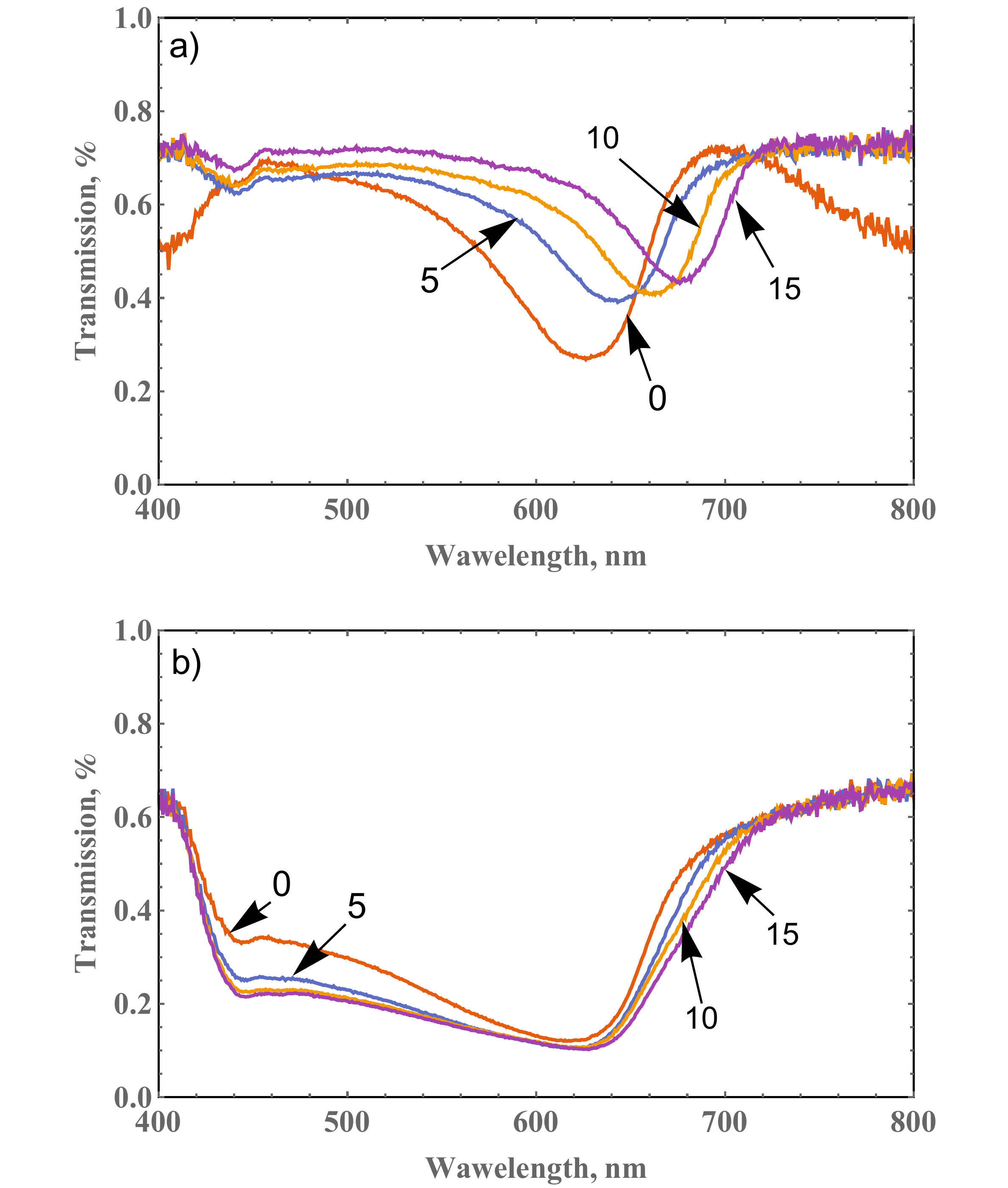}
\caption{
Spectral shift and narrowing of the selective reflection band under the action of ethanol for composition $\texttt{A}_{0.73} \texttt{B}_{0.23}$: a) sample thickness $35\; \mu m$  b) sample thickness is about $160 \;\mu m$. The numbers correspond to the expose time to ethanol in minutes.
}
\label{FIG:4ab}
\end{figure}
\begin{figure}[ht!]
\centering
\includegraphics[width=\columnwidth]{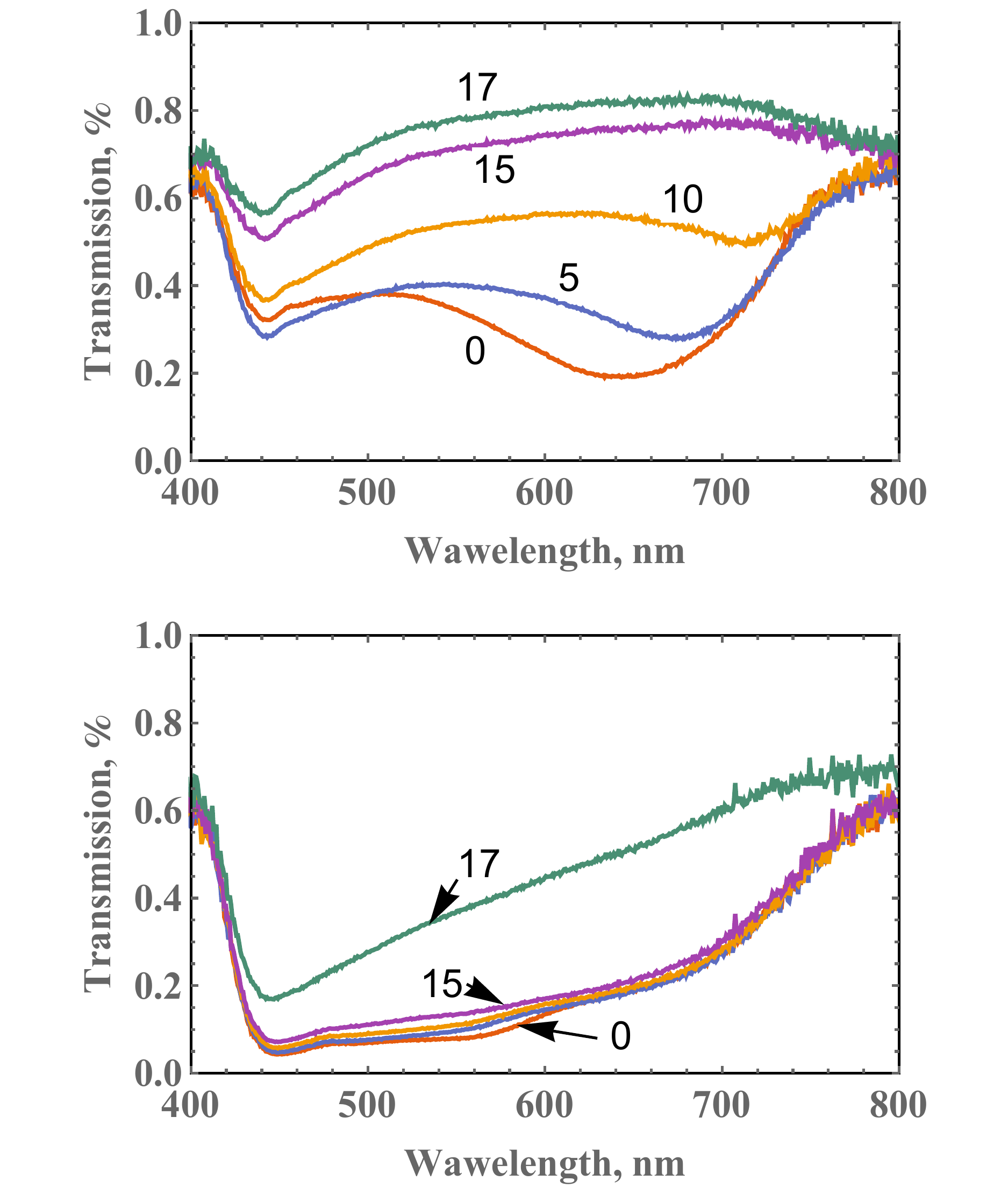}
\caption{
Spectral shift and narrowing of the selective reflection band under the action of toluene ( samples thickness is about 180 microns): a) for  composition $\left({\texttt{A}_{0.73} \texttt{B}_{0.23}}\right)_{0.9}\texttt{C}_{0.1}$ ( more viscous than $\texttt{A}_{0.73} \texttt{B}_{0.23}$), b) for composition $\left({\texttt{A}_{0.73} \texttt{B}_{0.23}}\right)_{0.8}\texttt{C}_{0.2}$ (more viscous than composition $\left({\texttt{A}_{0.73} \texttt{B}_{0.23}}\right)_{0.9}\texttt{C}_{0.1}$. The numbers correspond to the expose time to toluene in minutes. 
}
\label{FIG:6ab}
\end{figure}

The character of spectral response also allows identifying the particular VOC. Recording spectral changes (via Ocean Optics spectrometer) in thin films presenting individual compositions under the same experimental conditions related to VOC evaporation for the whole sensor allows detecting VOCs at much lower exposure times and calculating the shift of the selective reflection band and birefringence; this is illustrated in Fig. \ref{FIG:3} (a) for the pair ethanol + composition $\texttt{A}_{0.73} \texttt{B}_{0.23}$ and a film thickness of c.a. $35 \; \mu m$. The shift of the selective reflection band is already observed after 5 minutes of exposure and increases after 20 minutes; the width of the selective reflection band decreases as a function of exposure time. The refractive indices were measured by means of refractometry for composition $\texttt{A}_{0.73} \texttt{B}_{0.23}$ and turned out to be $1.485$ and $1.515$ at $21^\circ \;\texttt{C}$. Eq.\eqref{Eq:1}  were used to calculate helical pitch of the system $p=419\; \texttt{nm}$ by dividing a magnitude of the spectral position of the SRB center by the  average refractive index $n_{av}=1.505$. It can be seen that the width of the selective reflection band determined from Eq.\eqref{Eq:2} is smaller than the width found in experiments. This can be explained in terms of non-planar structure of CLC films, slight fluctuations of helical pitch, and collimation of a light beam used in experiments. Thicker samples display much wider SRBs since disorientation of the domains and light scattering contribute significantly to SRB widening. This is seen in Fig. \ref{FIG:3} b) where spectral changes for a thicker film (c.a. 160 microns) of composition $\texttt{A}_{0.73} \texttt{B}_{0.23}$ in ethanol atmosphere are presented.  The smaller spectral  shift of c.a. $25\; \texttt{nm}$ of the whole SRB towards longer wavelength is seen in the thicker sample comparing to a shift of about $50\; \texttt{nm}$ in the thinner sample (Fig. \ref{FIG:4ab} (a)). The similar spectroscopic changes occur for all other compositions and other VOCs. The spectral changes are all reversible, taking a shortest time for a pair composition $\texttt{A}_{0.73} \texttt{B}_{0.23}$ + ethanol and much longer time for compositions $\left({\texttt{A}_{0.73} \texttt{B}_{0.23}}\right)_{x}\texttt{C}_{1-x}$ and toluene. The role of Wacker compound in increasing viscosity of compositions and decreasing diffusion rate of toluene is seen in figures \ref{FIG:4ab}, \ref{FIG:6ab} where transmission spectra of the two films with similar thicknesses ( c.a. $180 \; \mu m$) and increasing concentration of Wacker oligomer are presented. It can be seen that complete isotropisation is achieved after 15 min in a film of composition $\texttt{A}_{0.73} \texttt{B}_{0.23}$ (Fig.\ref{FIG:4ab} (a)), it is not yet achieved in a film of composition $\left({\texttt{A}_{0.73} \texttt{B}_{0.23}}\right)_{0.9}\texttt{C}_{0.1}$ in which concentration of Wacker oligomer is close to $10\%$ after 17 min of exposure (Fig. \ref{FIG:6ab} (a)) and it is far from being achieved in a film of composition $\left({\texttt{A}_{0.73} \texttt{B}_{0.23}}\right)_{0.8}\texttt{C}_{0.2}$ after 25 min (Fig. \ref{FIG:6ab} (b)).
\par
Although the impressive selectivity of this sensor is demonstrated for the first time, the sensitivity of the sensor is somewhat depressed by non-planar structure of chiral droplets that increases with increasing the droplet’s size: it is well known that free cholesteric surface of large samples exposed to the air tend to adopt homeotropic orientation. This is seen in Fig.\ref{FIG:4ab}, where the transmission spectra show significant widening of the spectrum of composition $\texttt{A}_{0.73} \texttt{B}_{0.23}$ due to the disorientation of the domains and increased light scattering when the thickness of the sample increases by a factor of ten. A similar spectral widening is observed for other samples. Some degree of widening can be explained in terms of domains disorientation. The results of light transmission modeling conducted in accordance to Berreman method \cite{berreman1972optics} for composition $\texttt{A}_{0.73} \texttt{B}_{0.23}$ are the following. The refractive indices were measured to be $n_o= 1.485$   and $n_e=1.515$, the birefringence of the sample then is $\Delta n=0.03$, the order parameter was assumed to be $S=0.6$. The calculations were performed for the samples of thickness 12 helical pitches and 48 helical pitches ( the details of the calculations are presented in supplementary materials). Transmission spectra of thin samples can be well approximated by averaging over different orientations of the domains ranging from $0^\circ$ (planar orientation) to $50.3^\circ$ degrees. Transmission spectra of thicker samples can be approximated by averaging over mentioned domain orientations. Thus, the disorientation inherently presented in the droplets can be suppressed by decreasing the size of the droplet that should also increase the overall sensitivity of the method. 
%-------------------
\section{Conclusion}
Compositions of liquid crystals, liquid crystalline oligomers and non-mesogenic compound were designed and studied as the elements of LC-nose. The array of eight elements, liquid crystal compositions with varying viscosities and affinities to four VOCs ( ethanol, acetic acid, toluene and pyridine ), displayed a selective spectral  response (defined as a rate of color change in red and green channels of CCD sensor) to these VOCs. Decreasing the size of the elements increases the rate of response and therefore sensitivity of the response. The use of LC arrays for gas detection opens new opportunities in developing novel compact and wearable LC sensors.

\bibliography{aipsamp}
\newpage
%\begin{acknowledgments}
% Put your acknowledgments here.
%\end{acknowledgments}

\section{Supplementary information. }
Let us consider a mosaic cholesteric film consisting of individual identical domains randomly oriented with respect to the light source.
We shall define domain orientation via angle $\phi$ as shown in Fig.\ref{FIG1}. 
\begin{figure}[ht!]
\centering
\includegraphics[width=\columnwidth]{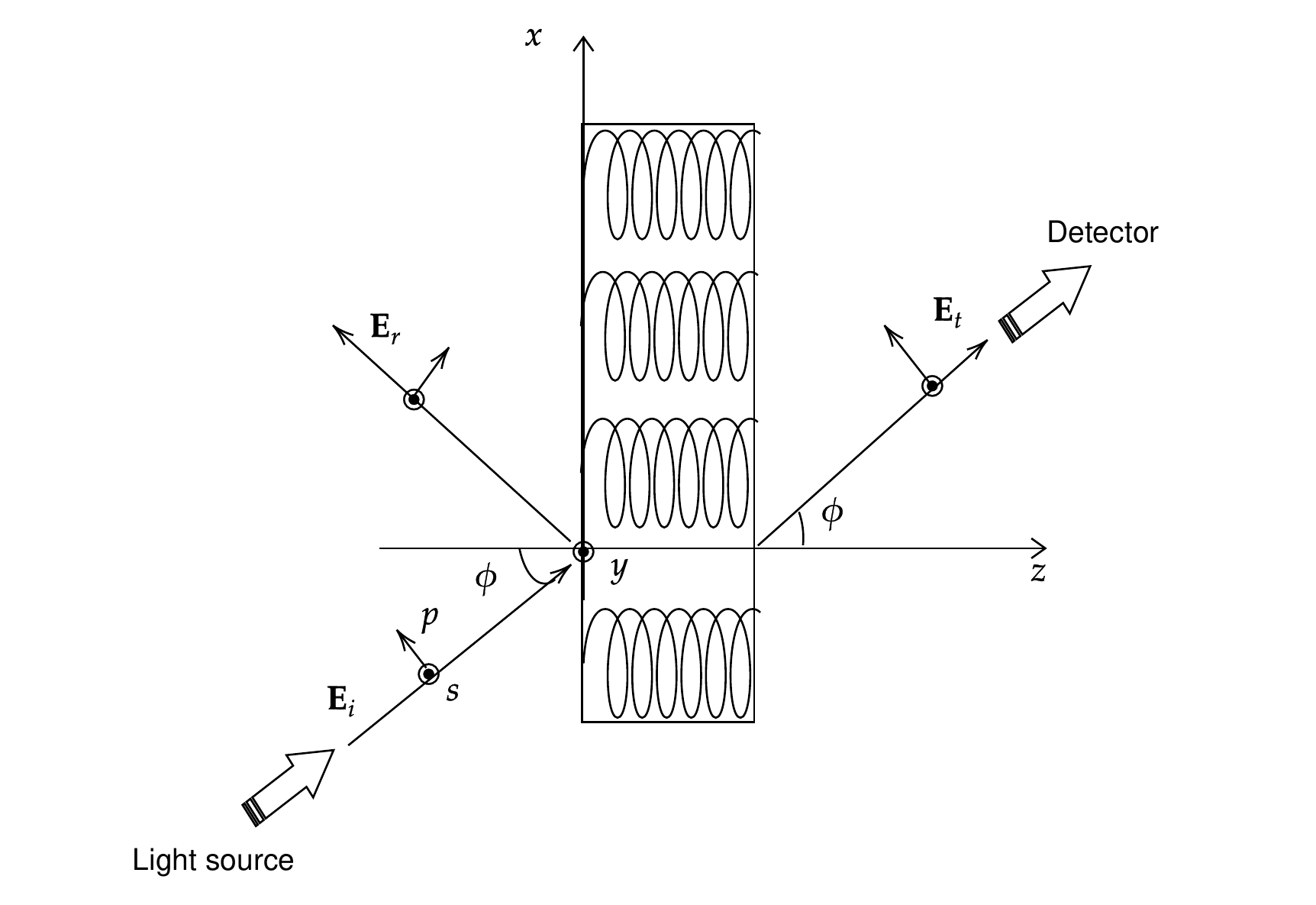}
\caption{Transmission through an individual domain of mosaic cholesteric.}
\label{FIG1}
\end{figure}
Each domain is surrounded by an isotropic (zero order parameter) media. In the local coordinate system the domain is described by the cholesteric LC model for the dielectric tensor within $0 \le z \le 1$: 
\begin{gather}
\hat{\epsilon} =
    \begin{pmatrix} 
    \bar{\epsilon}+\frac{1}{2}\delta \epsilon \cos\left({4 \pi z /p}\right)&
    \frac{1}{2}\delta \epsilon \sin\left({4 \pi z /p}\right) & 0
    \\ 
     \frac{1}{2}\delta \epsilon \sin\left({4 \pi z /p}\right)&
     \bar{\epsilon}-\frac{1}{2}\delta \epsilon \cos\left({4 \pi z /p}\right)&
     0
     \\
     0 & 0 & \bar{\epsilon}
    \end{pmatrix}\\
    \notag
    \bar{\epsilon} = \left({2 \epsilon_{0,xx}+ \epsilon_{0,yy}}\right)/3 = (n_e^2+n_o^2)/2;\\
    \notag
    \delta \epsilon = S \left({\epsilon_{0,xx} - \epsilon_{0,yy}}\right)= n_e^2-n_o^2
\end{gather}
, where $n_e, n_o$ are the extraordinary and ordinary refractive indices (see main text); 
$S$ is the order parameter related to the concentration of the absorbed gas; $p$ is the cholesteric pitch.
The director at the front face $z=0$ points in $x-$direction.
Outside of that region the dielectric constant is $\bar\epsilon$, so that the sample becomes fully transparent in absence of the order parameter.
\par
The EM field in $0 \le z \le 1$ is governed by Berreman equation:
\begin{gather}
\label{EQ:2}
    \frac{\partial \boldsymbol{\psi}}{\partial z}= \frac{2 \pi}{\lambda} \hat{\mathcal{D}} \boldsymbol{\psi}\\
    \notag
 \boldsymbol{\psi}=
    \begin{pmatrix}
    E_x\\
    i H_y\\
    E_y\\
    -i H_x
    \end{pmatrix}\\
    \notag
    \hat{\mathcal{D}} =
    \begin{pmatrix}
    0 & 1-X^2/\epsilon_{zz} & 0 & 0 \\
    -\epsilon_{xx} & 0 & -\epsilon_{xy} & 0\\
    0 & 0 & 0 & 1\\
    -\epsilon_{xy} & 0 & X^2-\epsilon_{yy} & 0
    \end{pmatrix}
\end{gather}
, where $\lambda$ is the wavelength; $X = k_x \lambda/ 2 \pi = n_{av} \sin \phi$ and $\phi$ is the reflection/transmission angle (See Fig.\ref{FIG1}); $n_{av} = \sqrt{\bar\epsilon}$ is the refractive index of the surrounding media. 
\par
The formal solution of Eq.\eqref{EQ:2} is:
\begin{equation}
\label{EQ:3}
    \boldsymbol{\psi}\left({z}\right)= \texttt{Exp}\left({\frac{2 \pi}{\lambda}\int \limits_{1}^{z} \hat{\mathcal{D}}dz }\right) \boldsymbol{\psi}\left(1_{+}\right)
\end{equation}
Using the trapezoidal approximation for the integral we obtain the following expression for the propagator $\hat{\mathcal{P}}$:
\begin{gather}
\label{EQ:4}
\boldsymbol{\psi}\left(0_{-}\right) = \hat{\mathcal{P}}\boldsymbol{\psi}\left(1_{+}\right);\\
\notag
\hat{\mathcal{P}} \approx \texttt{Exp}\left({\frac{\pi}{\lambda}\hat{\mathcal{D}}\left({0}\right) \delta z}\right)
\\
\notag
\times
\left({\prod \limits_{j=1}^{N_{0}-1}  \texttt{Exp}\left({\frac{ 2 \pi}{\lambda}\hat{\mathcal{D}}\left({j \delta z}\right) \delta z}\right)}\right)
 \texttt{Exp}\left({\frac{\pi}{\lambda}\hat{\mathcal{D}}\left({1}\right) \delta z}\right)
\end{gather}
, here $\delta z = 1/ \left({p N_0}\right)$ and $N_0$ is the finesse of the numerical approximation.
\par
On the front half space we have
\begin{gather}
\label{EQ:5}
 \boldsymbol{\psi}\left(0_{-}\right) = 
 \begin{pmatrix}
 \cos \left({\phi}\right) E_{ip}\\
  i n E_{ip}\\
 E_{is}\\
 i n \cos \left({\phi}\right) E_{is}
 \end{pmatrix}
 +
 \begin{pmatrix}
 \cos \left({\phi}\right) E_{rp}\\
 - i n E_{rp}\\
 E_{rs}\\
 -i n \cos \left({\phi}\right) E_{rs}
 \end{pmatrix}
 = \\
 \notag
 =\hat{J}_{0-}
 \begin{pmatrix}
 E_{is}\\
 E_{rs}\\
 E_{ip}\\
 E_{rp}
 \end{pmatrix};
 \\
 \notag
 \hat{J}_{0-} =
 \begin{pmatrix}
 0 & 0 & \cos \left({\phi}\right) & \cos \left({\phi}\right)\\
 0 & 0 & i n & - i n\\
 1 & 1 & 0 & 0\\
 i n  \cos \left({\phi}\right) & -i n  \cos \left({\phi}\right) & 0 & 0
 \end{pmatrix}
\end{gather}
On the back half-space we have:
\begin{gather}
\label{EQ:6}
 \boldsymbol{\psi}\left(1_{+}\right) = 
 \begin{pmatrix}
 \cos \left({\phi}\right) E_{tp}\\
  i n E_{tp}\\
 E_{ts}\\
 i n \cos \left({\phi}\right) E_{ts}
 \end{pmatrix}
 = \hat{J}_{1+} 
 \begin{pmatrix}
 E_{ts}\\
 0\\
 E_{tp}\\
 0
 \end{pmatrix};
 \\
 \notag
 \hat{J}_{1+} =
 \begin{pmatrix}
 0 & 0 & \cos \left({\phi}\right) & 0\\
 0 & 0 & i n & 0\\
 1 & 0 & 0 & 0\\
 i n  \cos \left({\phi}\right) & 0 & 0 & 0
 \end{pmatrix}
\end{gather}
Here $\mathbf{E}_{i}, \mathbf{E}_{r}, \mathbf{E}_{t}$ are the incoming, reflected and transmitted electric field respectively.
\par
Putting Eq.\eqref{EQ:5}, \eqref{EQ:6} into Eq.\eqref{EQ:4} we obtain the transfer matrix $\hat{T}$ as:
\begin{gather}
\label{EQ:7}
 \begin{pmatrix}
 E_{is}\\
 E_{rs}\\
 E_{ip}\\
 E_{rp}
 \end{pmatrix} =
 \hat{T}
 \begin{pmatrix}
 E_{ts}\\
 0\\
 E_{tp}\\
 0
 \end{pmatrix};\\
 \notag
 \hat{T} = \hat{J}^{-1}_{0-} \hat{\mathcal{P}} \hat{J}_{1+}
\end{gather}
As a sanity check we get the transfer matrix for the infinitely thin sample $\hat{\mathcal{P}} =\hat{1}$ obtaining:
\begin{equation*}
    \hat{T} = 
    \begin{pmatrix}
    1 & 0 & 0 & 0\\
    0 & 0 & 0 & 0\\
    0 & 0 & 1 & 0\\
    0 & 0 & 0 & 0
    \end{pmatrix}
\end{equation*}

The solution of Eqs.\eqref{EQ:7} can be written as:
\begin{gather}
\label{EQ:8}
    \begin{pmatrix}
    E_{tp} \\
    E_{ts}
    \end{pmatrix}
    = \begin{pmatrix}
    t_{pp} & t_{ps}\\
    t_{sp} & t_{ss}
    \end{pmatrix}
    \begin{pmatrix}
    E_{ip}\\
    E_{is}
    \end{pmatrix}
    = \\
    \notag
    =
    \begin{pmatrix}
    T_{33} & T_{31}\\
    T_{13} & T_{11}
    \end{pmatrix}^{-1}
     \begin{pmatrix}
    E_{ip}\\
    E_{is}
    \end{pmatrix};
    \\
    \notag
    \begin{pmatrix}
    E_{rp} \\
    E_{rs}
    \end{pmatrix}
    = \begin{pmatrix}
    r_{pp} & r_{ps}\\
    r_{sp} & r_{ss}
    \end{pmatrix}
    \begin{pmatrix}
    E_{ip}\\
    E_{is}
    \end{pmatrix}
    = \\
    \notag
    =
    \begin{pmatrix}
    T_{43} & T_{41}\\
    T_{23} & T_{21}
    \end{pmatrix}
    \begin{pmatrix}
    T_{33} & T_{31}\\
    T_{13} & T_{11}
    \end{pmatrix}^{-1}
     \begin{pmatrix}
    E_{ip}\\
    E_{is}
    \end{pmatrix}
\end{gather}
Here the input fields are:
\begin{gather}
\label{EQ:9}
    \mathbf{E}_{ip} =
    \begin{pmatrix}
    1\\
    0
    \end{pmatrix};\\
    \notag
    \mathbf{E}_{is} =
    \begin{pmatrix}
    0\\
    1
    \end{pmatrix}
\end{gather}
We can change the basis to that of left/right polarizations as:
\begin{gather}
    \begin{pmatrix}
    E_{ip}\\
    E_{is}
    \end{pmatrix}
    =\hat{C}_i
    \begin{pmatrix}
    E_{iL}\\
    E_{iR}
    \end{pmatrix}
    = \frac{1}{\sqrt{2}}
    \begin{pmatrix}
    1 & 1\\
    i & -i 
    \end{pmatrix}
    \begin{pmatrix}
    E_{iL}\\
    E_{iR}
    \end{pmatrix};\\
    \notag
    \begin{pmatrix}
    E_{rp}\\
    E_{rs}
    \end{pmatrix}
    =\hat{C}_r
    \begin{pmatrix}
    E_{rL}\\
    E_{rR}
    \end{pmatrix}
    = \frac{1}{\sqrt{2}}
    \begin{pmatrix}
    1 & 1\\
    -i & i 
    \end{pmatrix}
    \begin{pmatrix}
    E_{rL}\\
    E_{rR}
    \end{pmatrix};
\end{gather}
In the L/R basis Eq.\eqref{EQ:8} becomes:
\begin{gather}
\label{EQ:11}
    \begin{pmatrix}
    E_{tL} \\
    E_{tR}
    \end{pmatrix}
    = \begin{pmatrix}
    t_{LL} & t_{LR}\\
    t_{RL} & t_{RR}
    \end{pmatrix}
    \begin{pmatrix}
    E_{iL}\\
    E_{iR}
    \end{pmatrix}
    = \hat{C}_i^{-1} \hat{t}_{ps} \hat{C}_i
   \begin{pmatrix}
    E_{iL}\\
    E_{iR}
    \end{pmatrix};
    \\
    \notag
    \begin{pmatrix}
    E_{rL} \\
    E_{rR}
    \end{pmatrix}
    = \begin{pmatrix}
    r_{LL} & r_{LR}\\
    r_{RL} & r_{RR}
    \end{pmatrix}
    \begin{pmatrix}
    E_{iL}\\
    E_{iR}
    \end{pmatrix}
    = 
    \hat{C}_r^{-1} \hat{t}_{ps} \hat{C}_i
     \begin{pmatrix}
    E_{iL}\\
    E_{iR}
    \end{pmatrix}
\end{gather}
applied to
\begin{gather}
\label{EQ:9}
    \mathbf{E}_{iL} = \frac{1}{\sqrt{2}}
    \begin{pmatrix}
    1\\
    -i
    \end{pmatrix};\\
    \notag
    \mathbf{E}_{iR} = \frac{1}{\sqrt{2}}
    \begin{pmatrix}
    1\\
    i
    \end{pmatrix}
\end{gather}
The observable (reflection or transmission) is given by the square of absolute value of the left hand side of either Eq.\eqref{EQ:8} or Eq.\eqref{EQ:11}.
For un-polarized light incident on the domain the transmission/reflection is given by:
\begin{gather}
\label{EQ:13}
    T\left({\lambda, \phi}\right) = 
    \frac{1}{2}\left[{t_{pp}^2+t_{ps}^2+t_{sp}^2+t_{ss}^2}\right]\\
    \notag
    R\left({\lambda, \phi}\right) = 
    \frac{1}{2}\left[{r_{pp}^2+r_{ps}^2+r_{sp}^2+r_{ss}^2}\right]
\end{gather}
Regardless of the input field the eigenvalues of $\hat{T}\left({\lambda,\phi}\right)$ plays a crucial role in determining stop band.
Namely if one pair of those eigenvalues becomes purely imaginary. 
Instead of relying on the eigenvalue decomposition it is possible to use Eqs.\eqref{EQ:13} for the same purpose.
Indeed, let us first fix the domain orientation at $\phi =0$ and assume $p \ll 1$. In the stop band a circular polarization (in our case L polarized signal since the director rotates R when it goes from left to right half space) is getting totally reflected. If we assume constant value of $S(z)$, the reflection occurs at $\sqrt{\bar{\epsilon}-\delta \epsilon/2} \leq \lambda/p \leq \sqrt{\bar{\epsilon}+\delta \epsilon/2}$ with the band center given by $\lambda_c/p = \sqrt{\bar{\epsilon}}$. For an arbitrary domain orientation the stop band center can be obtained numerically by finding maximum of $R$ in Eq.\eqref{EQ:13} upon fixed angle of incidence. Alternatively it can be estimated from the Bragg condition:
\begin{equation}
\label{EQ:14}
    \frac{\lambda_{c}}{p}=\sqrt{\bar{\epsilon}}\cos{\left({\phi}\right)}
\end{equation}.
For our simulations parameters the above condition turned out to be precise. For our multi-domain sample individual stop-bands provide little information and it is advisable to look at a relative spectral weight of individual domains. We shall define it as amount of energy reflected by the domain normalized by the energy reflected by the normally oriented domain ($\phi =0$):
\begin{equation}
\label{EQ:15}
    W\left({\phi}\right) = \frac{\int d\lambda R\left({\lambda,\phi}\right)}{\int d\lambda R\left({\lambda,0}\right)}
\end{equation}
For our numerical simulations we focus on Composition $\texttt{A}_{0.73}\texttt{B}_{0.23}$ (See Fig.4 of the main text).
The experiment suggests the following bandwidth $430 \leq \lambda \leq 675 \; \texttt{nm}$. 
Measured refractive indices are $n_e = 1.515$ and $n_o = 1.515$ yielding $p=675/\sqrt{\bar\epsilon} = 450 \; \texttt{nm}$ and $\phi_{max} = 50.3^\circ$. Mutual scattering events between domains are neglected. 
The results of the simulations are shown in Fig.\ref{FIG1} and Fig.\ref{FIG2}. They suggest that for wider samples $L/p>40$ the spectral weights of the domains become identical resulting in the wide band transmission spectra. Thin films favour $\phi=0$ normal domain orientations, with notable exception of viscous compositions (See Fig.6 main text). 
\par
The diffusion of toluene occurs more readily into tilted domains, thus reducing their order parameter and therefore spectral weight (See Fig.5).
In viscous compositions the opposite effect manifests itself. 
\begin{figure}[ht!]
\centering
\includegraphics[width=\columnwidth]{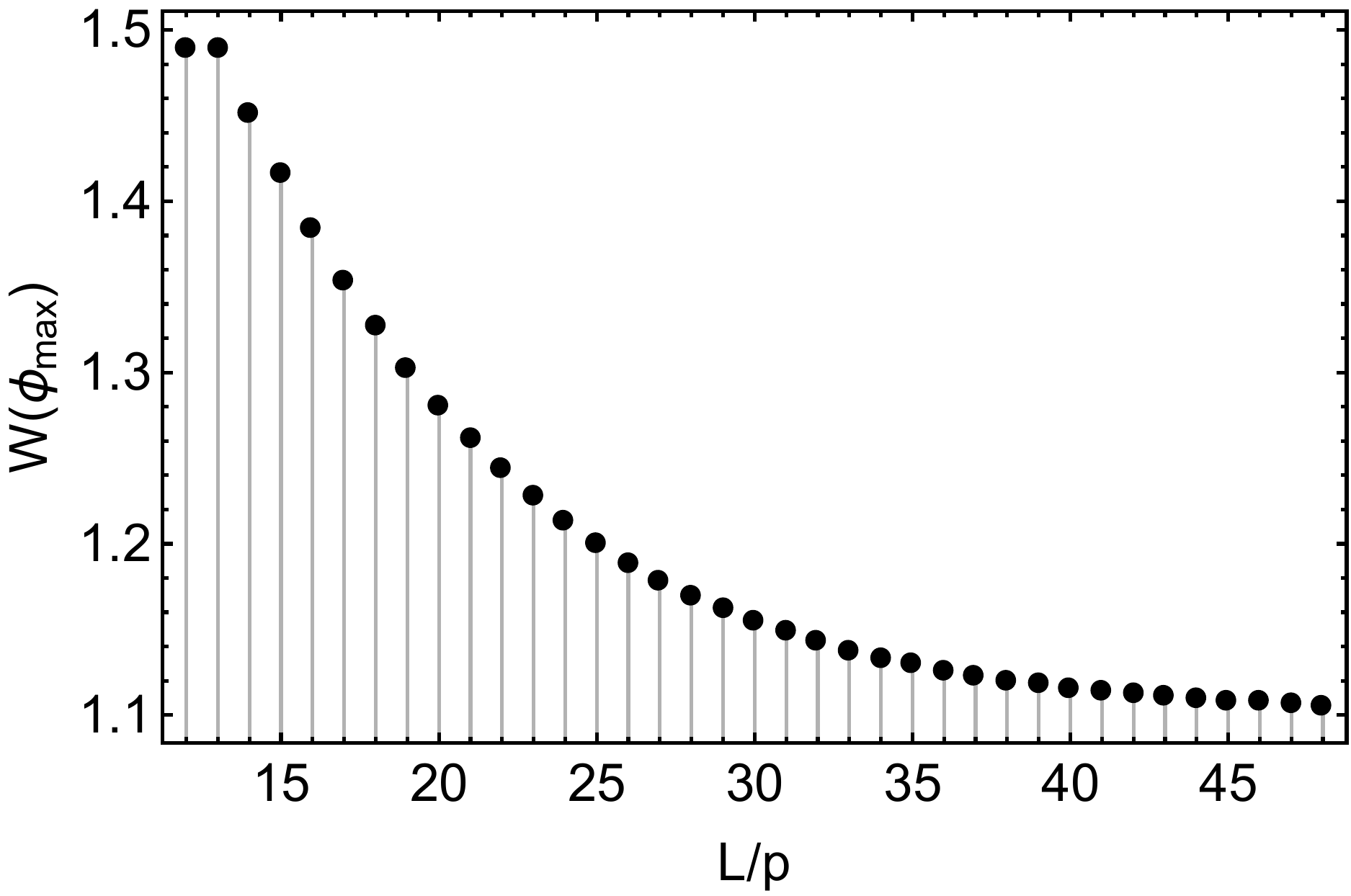}
\caption{Relative spectral weight for the domain oriented at $\phi_{max}=50.3^{\circ}$}
\label{FIG1}
\end{figure}

\begin{figure}[ht!]
\centering
\includegraphics[width=\columnwidth]{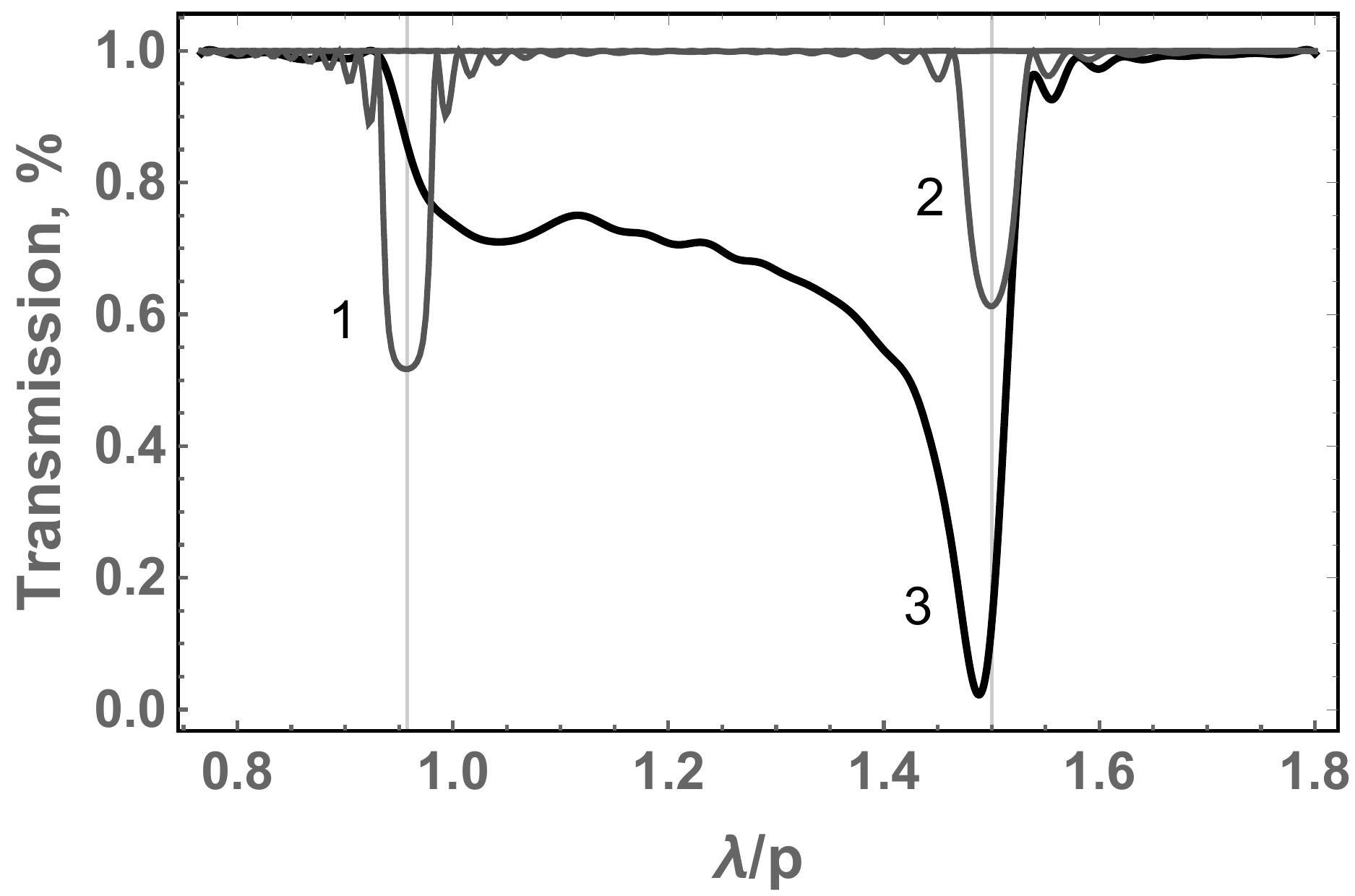}
\caption{Simulated transmission spectra for $L = 22 p$  individual domains contributions: 1) $\phi=50.34^\circ$ and 2) $\phi=0.0^\circ$.
3) is the transmission through uniformly distributed set of identical domains inclined by $-50.32^\circ \leq \phi \leq 50.32^\circ$. Vertical lines correspond to the Bragg conditions for the band center in Eq.\eqref{EQ:14}. The same brag condition assures that for uniform distribution primary contribution to the spectra is due the normally oriented domains.}
\label{FIG2}
\end{figure}

\end{document}